\pdfoutput=1
\documentclass[aps,prl,reprint,groupedaddress]{revtex4-1}
\usepackage{graphicx}
\usepackage{amssymb,amsfonts,amsmath}
\usepackage{epsfig}
\usepackage{color}
\usepackage[normalem]{ulem}
\def\lsim{\mathrel{\rlap{\lower3pt\hbox{\hskip1pt$\sim$}}
     \raise1pt\hbox{$<$}}}
\def\gsim{\mathrel{\rlap{\lower3pt\hbox{\hskip1pt$\sim$}}
     \raise1pt\hbox{$>$}}}
\newcommand{\beq}{\begin{equation}}
\newcommand{\eeq}{\end{equation}}
\newcommand{\bea}{\begin{eqnarray}}
\newcommand{\eea}{\end{eqnarray}}

\begin{document}
\title{Is there a flavor hierarchy in the deconfinement transition of QCD?}
\author{Rene Bellwied$^a$, Szabolcs Borsanyi$^b$, Zoltan Fodor$^{b,c,d}$, S\'andor D. Katz$^{c,e}$, Claudia Ratti$^f$ \\
$^a$ \small{\it University of Houston, Houston, TX 77204, USA}\\
$^b$ \small{\it Department of Physics, Wuppertal University, Germany}\\
$^c$ \small{\it E\"otv\"os University, Theoretical Physics, P\'azm\'any P. S 1/A, H-1117, Budapest, Hungary}\\
$^d$ \small{\it J\"ulich Supercomputing Centre, Forschungszentrum J\"ulich, D-52425 J\"ulich, Germany}\\
$^e$ \small{\it MTA-ELTE Lend\"ulet Lattice Gauge Theory Research Group, Budapest, Hungary}\\
$^f$ \small{\it Universit\`a degli Studi di Torino and INFN, Sezione di Torino, via Giuria 1, I-10125 Torino, Italy}\\
}

\begin{abstract}
We present possible indications for flavor separation during the QCD crossover
transition based on continuum extrapolated lattice QCD calculations of
higher order susceptibilities. We base our findings on flavor specific
quantities in the light and strange quark sector. We propose a possible experimental verification of our prediction, based on the measurement of
higher order moments of identified particle multiplicities. Since all our
calculations are performed at zero baryochemical potential, these results are
of particular relevance for the heavy ion program at the LHC.  \end{abstract}
\maketitle

It is expected that the Universe, only a few microseconds after the Big Bang, underwent a transition in which matter converted from a state of free quarks and gluons to a state of colour-neutral particles of finite mass, namely the hadrons, which populate the Universe today. This
transition can be reproduced in the laboratory, in relativistic heavy-ion collisions currently taking place at the Large Hadron Collider (LHC) at CERN and the RHIC facility at Brookhaven National Laboratory (BNL).

From the theoretical point of view, this transition can be studied from first
principles, by simulating QCD on a discretized lattice. Results of such studies
show that the finite-temperature QCD transition is merely an analytic crossover
\cite{Aoki:2006we}. This means that during the cooling of the Universe the
system transitioned from the phase dominated by coloured particles to the
hadronic phase over an extended period of time without the emission of
latent heat. Since no unambiguous temperature can be assigned to this
transition, the question arises whether hadrons of different quark composition
freeze-out simultaneously or exhibit a flavor hierarchy \cite{Ratti:2011au}.
This question is relevant since the reported strangeness enhancement at SPS,
RHIC and LHC energies (in particular in the multi-strange particle sector
\cite{Hippolyte:2012np,Blume:2011sb,Agakishiev:2011ar}), and the discovery of
hypernuclei formation at RHIC \cite{Abelev:2010rv} suggest the possibility of
increased strange bound state production at the highest available collision
energies.  Furthermore, recent measurements in relativistic heavy-ion
collisions at the LHC indicate a separation of chemical freeze-out temperatures
between light and strange quark hadrons
\cite{Preghenella:2011np,Abelev:2012wca}.

In the present paper we show a set of observables, obtained by means of
continuum-extrapolated lattice QCD simulations with physical masses, which
indicate a flavor separation in the transition region of QCD. Such quantities
are based on flavor-specific fluctuations, as well as on correlations between
different flavors or conserved charges.

In a series of papers, the Wuppertal-Budapest Collaboration used the tree level
improved Symanzik gauge action and a staggered fermionic action with 2-level
stout improvement (for a precise definition of the action see
\cite{Aoki:2005vt}). One of the advantageous features of stout smearing is the
improvement for the pion mass splitting,
typical for staggered QCD simulations.
In this paper we focus on observables that are not sensitive to valence pions
and remove the pion mass splitting effect in the sea by making a continuum
extrapolation.

To carry out the continuum extrapolation at a given temperature $T$ we
simulate for several $N_t$ values, or lattice spacings: $N_t$=6,8,10,12.
At $T_c$ these temporal extensions correspond to
$a=$0.22, 0.16, 0.13 and 0.11 fm lattice spacings,
respectively. In order to keep the same physical content when we decrease
$a$, the kaon decay constant and the kaon mass were tuned to
their physical values and we used for the strange to light quark mass ratio
also its physical value ($m_s$/$m_{ud}\approx 28$) determined in
Ref. \cite{Aoki:2009sc}. Lattice simulations are always carried
out in a finite volume.  In our case the
relevant observables (baryon, strange and light quark number
fluctuations) are mostly carried by the kaons and protons in the hadronic
phase. Pions which are mostly sensitive to
finite volume effects do not carry a net light quark number.
Therefore, we expect that finite volume effects of the present
analysis are small and are hidden by the statistical errors.

Fluctuations of conserved charges can be expressed in a grand canonical
ensemble as the derivatives of the partition function with respect to the
conserved charge chemical potential.
In QCD the net $u$, $d$ and $s$ quark numbers are conserved:
we introduce $\mu_u$, $\mu_d$ and $\mu_s$ as the corresponding chemical
potentials.
Fluctuations are then expressed in terms of
derivatives of the pressure $p$ of the equilibrated system:
\beq
\chi^{uds}_{lmn}=\frac{\partial^{l+m+n}(p/T^4)}{\partial(\mu_u/T)^l \partial(\mu_d/T)^m\partial(\mu_s/T)^n}.
\eeq
Odd $l$+$m$+$n$ combinations are sensitive to non-vanishing chemical
potentials, but since our main interest in this paper is the physics
at LHC, we work with vanishing $\mu$-s. More precisely we
concentrate on some quadratic and quartic fluctuations (thus $l+m+n$=2 or 4)
and their ratios. Since these fluctuations are directly related to conserved
currents, no renormalization ambiguity should appear.

%
%

Our focus is on $\mu_L$ and $\mu_S$ which couple to
the net light flavor density and the strangeness density, respectively.
To study correlations with the flavor-mixed baryon number we also
introduce the respective chemical potential $\mu_B$:
\begin{eqnarray}
\frac{\partial}{\partial\mu_L}&=&
\frac12\frac{\partial}{\partial\mu_u}
+\frac12\frac{\partial}{\partial\mu_d}\,;\quad
\frac{\partial}{\partial\mu_S}=
-\frac{\partial}{\partial\mu_s}\nonumber\\
\frac{\partial}{\partial\mu_B}&=&
\frac13\frac{\partial}{\partial\mu_u}
+\frac13\frac{\partial}{\partial\mu_d}
+\frac13\frac{\partial}{\partial\mu_s}
\end{eqnarray}
The method for extracting the diagonal and off-diagonal quark number susceptibilities as well as the second and fourth order derivatives has been worked out in detail in Refs.~\cite{Gottlieb:1987ac,Allton:2002ziandAllton:2003vx,Borsanyi:2011sw,Bazavov:2012jq}.

\begin{figure}[h!]
\centerline{
\includegraphics*[width=6.2cm]{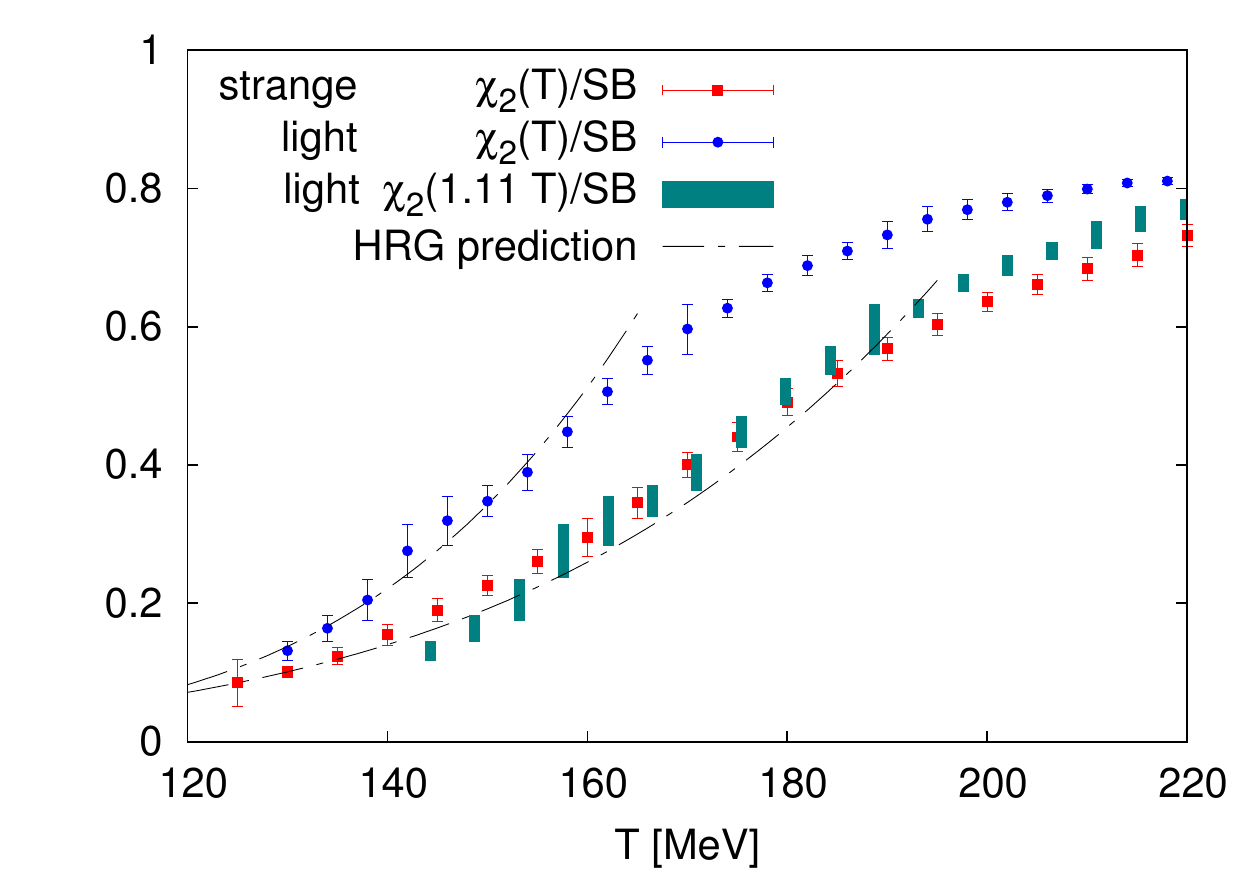}
}
\caption{\label{rescale}
Light and strange quark susceptibilities in the continuum limit (plotted as blue circles and red squares, respectively).
The transition temperatures defined by the inflection points for
$\chi_2^{L}$ ($150$~MeV) and for $\chi_2^s$ ($165$~MeV) differ by $\approx15$~MeV. A rescaling transformation is shown with bars.}
\end{figure}

Already in 2006 the Wuppertal-Budapest Collaboration showed that the
characteristic temperature of the transition in the strange sector (based on
$\chi^s_2$) is about 20~MeV higher than that for the light quark sector (based
on the chiral susceptibility \cite{Aoki:2006br} or $\chi^u_2$ in Ref.
\cite{Borsanyi:2010bp}). In order to understand these differences we determined
$\chi^u_2$ and $\chi^s_2$ with unprecedented accuracy \cite{Borsanyi:2011sw}.
Fig.~\ref{rescale} shows the continuum extrapolated $T$-dependence of the light
quark $\chi_2^{L}(T)$ and strange quark $\chi_2^{s}(T)$ susceptibilities. One
striking observation is an approximate scaling relation between the
$T$-dependencies of the light and strange quark susceptibilities, respectively.
A rescaling in $T$ for one of these observables closely reproduces the
other one: $\chi_2^{L}(T\cdot x)$=$\chi_2^{s}(T)$. The figure also shows this
rescaling relation, for which the rescaling factor x=1.11 is preferred. The
most important message of this relationship can be summarized as follows:
independently of a chosen characteristic point in the crossover region (e.g.
inflection point, halving point, or any other) and its relation to 
some physically motivated definition of the
transition temperature, the similarity transformation between the two curves
leads to a quite precise prediction for the difference between these transition
temperatures of light and strange quarks. We find that the 
characteristic temperature defined e.g. by the inflection point is 
$(x-1)\cdot T_c \simeq15$ MeV higher for the strange quark than for the light quarks.
(Here $T_c$ is the characteristic temperature for the
transition in the light quark sector). 

A model-dependent but enlightening approach to locate $T_c$
is to compare lattice data to the Hadron Resonance Gas
(HRG) model prediction. For the data in Fig.~\ref{rescale}, the highest temperature
of agreement with the HRG result is ambiguous and may depend on the
number of resonances included in the HRG partition sum. Therefore, we
continue our discussion with higher order fluctuations where
the HRG prediction is more robust and the point where lattice and
HRG results start to deviate is also a characteristic point of
the data set.

A recent work \cite{Bazavov:2013dta} suggested two interesting
susceptibility combinations, $v_1$ and $v_2$, which vanish in the
hadronic phase and become non-zero as soon as $s$ quark degrees of freedom
start to be liberated in the system.
We generalize these expressions to any flavor $f$:
\bea
v_1^f&=&\chi_{11}^{Bf}-\chi_{31}^{Bf};
\nonumber\\
v_2^f&=&\frac13\left(\chi_2^f-\chi_4^f\right)+2\chi_{13}^{Bf}-4\chi_{22}^{Bf}+2\chi_{31}^{Bf} \nonumber
\eea
Based on these combinations, it was shown
that strange quarks start to be deconfined at $T$$\gtrsim$157~MeV. Here we
compare, for the first time, such parameters for strange and light quarks.
We supplement this information with a third combination:
\bea
w^f&=&\chi_{13}^{Bf}-\chi_{11}^{Bf}\,.
\eea
This is more sensitive to the flavor content based on the higher
order of quark derivatives with respect to the baryon derivatives: in
particular, in the hadronic phase it only receives contributions from hadrons
containing more than one quark of flavor $f$.
In the following we consider $f=u,~s$ or $L$.

\begin{figure}[h!]
\includegraphics*[width=6.2cm]{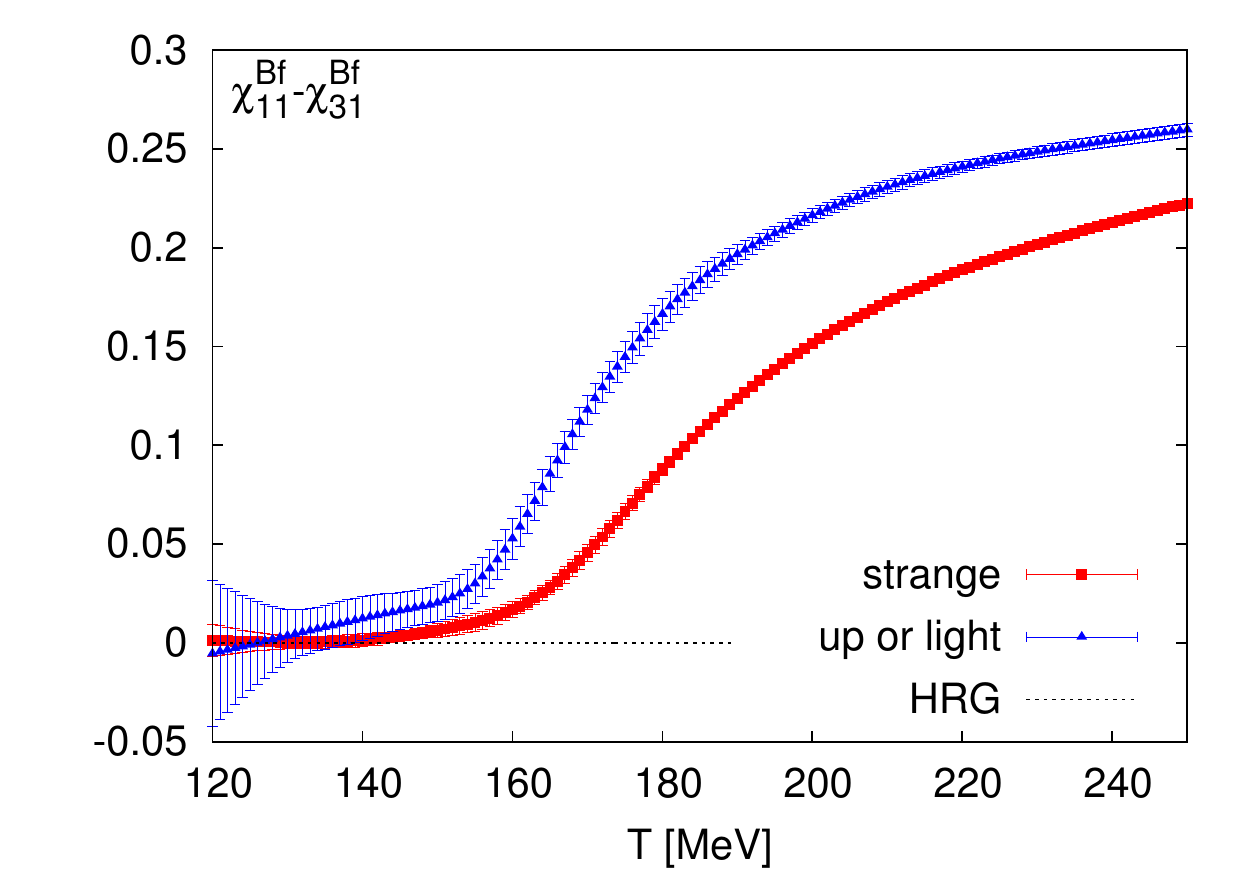}\\
\includegraphics*[width=6.2cm]{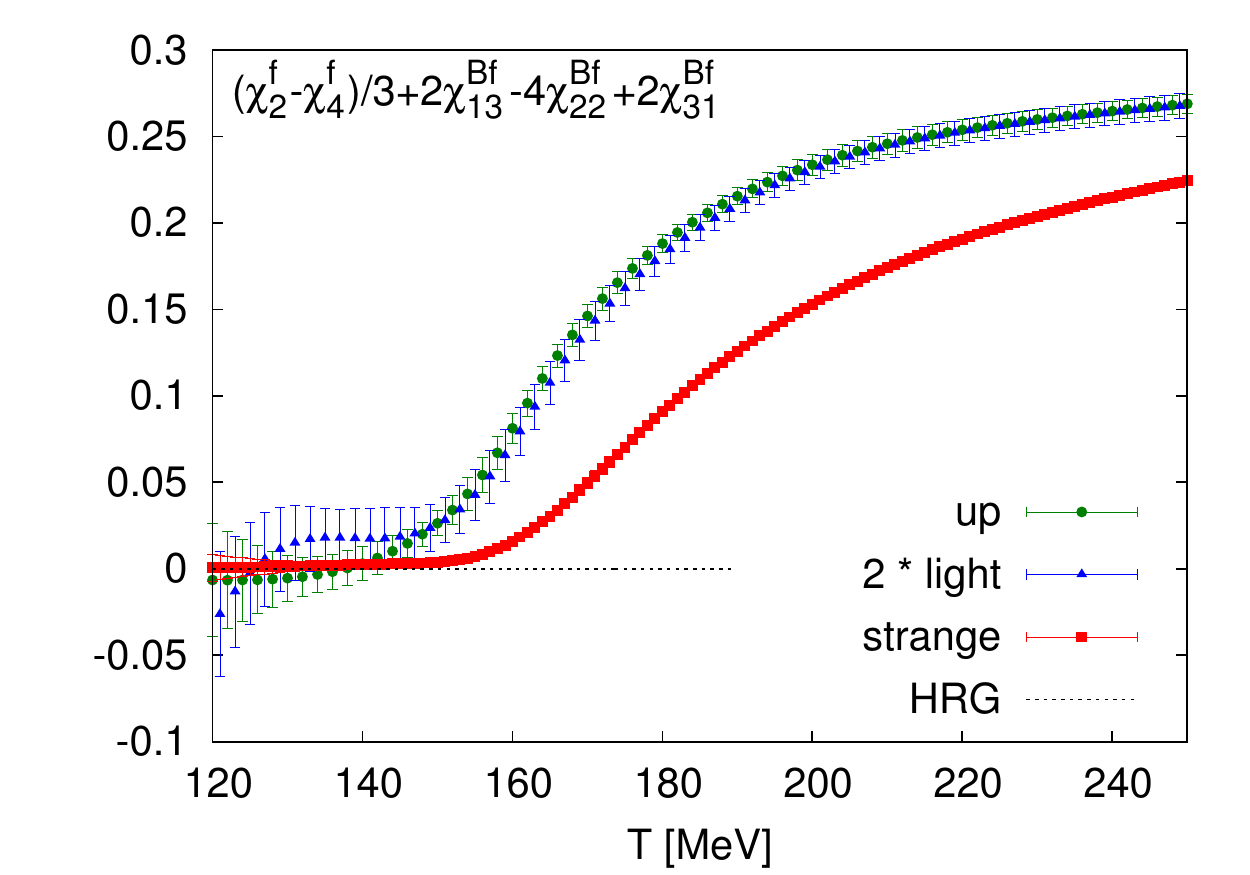}\\
\includegraphics*[width=6.2cm]{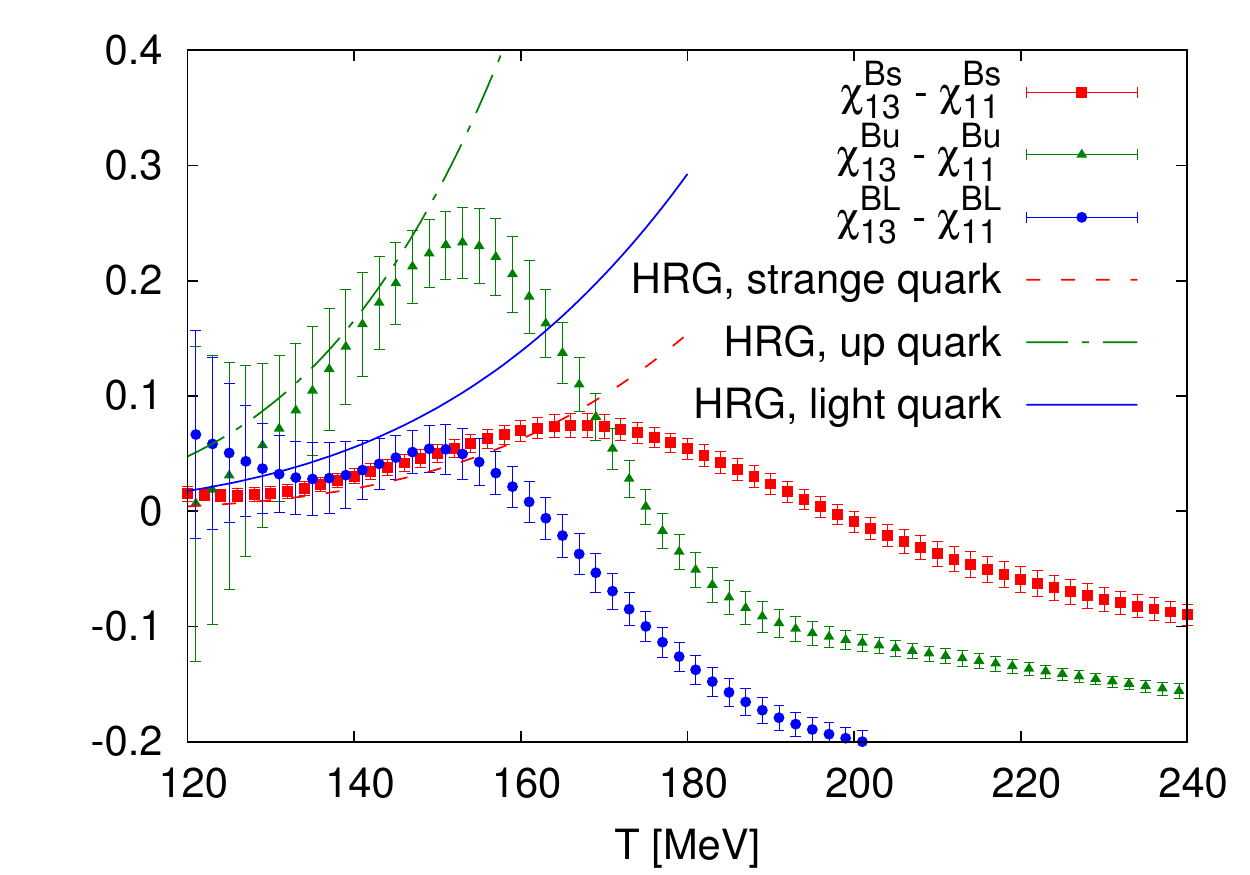}
\caption{\label{flavor-comb} The continuum extrapolated 
temperature dependence of the $v_1$,$v_2$,$w$ susceptibility
combinations for light and strange quarks in comparison to HRG calculations.
}
\end{figure}

Fig.~\ref{flavor-comb} shows our continuum extrapolated results in comparison
to HRG calculations.  Out of the three shown
observables, $w$ shows the strongest flavor separation. In all cases, a
deviation from the HRG at a certain minimum temperature can be
considered as the onset of the liberation of quarks of a given flavor. The
advantage of the HRG being
strictly zero for the first two derivatives is balanced by rather large error
bars in the lattice results, whereas in the case of $w$ the lattice calculation
shows a clearly identifiable characteristic peak at the temperature where it
starts to deviate from the HRG result. Notice that the temperatures at which
$v_1^f$ and $v_2^f$ deviate from the HRG model are lower than the inflection
points extracted from
$\chi_2^f$: this reflects the fact that the inflection point of $\chi_2^f$
defines the ``steepest" point of the transition, while $v_1^f$ and
$v_2^f$ deviate from the HRG model as soon as the first deconfined quark of
flavor $f$ appears.
For observables that are most sensitive to multi-strange content we see a
pronouncedly higher temperature of deviation from HRG  than for the analogous
quantity in the light sector. The contribution of multi-strange hadrons is
enhanced in a combination with higher $\frac{\partial}{\partial \mu_s}$
derivatives, like $\chi_2^s$ or $w$, whereas $v_1$ and $v_2$ signal the
liberated strangeness from all strange hadrons equally.

\begin{figure}[t]
\centerline{
\includegraphics*[width=6.2cm]{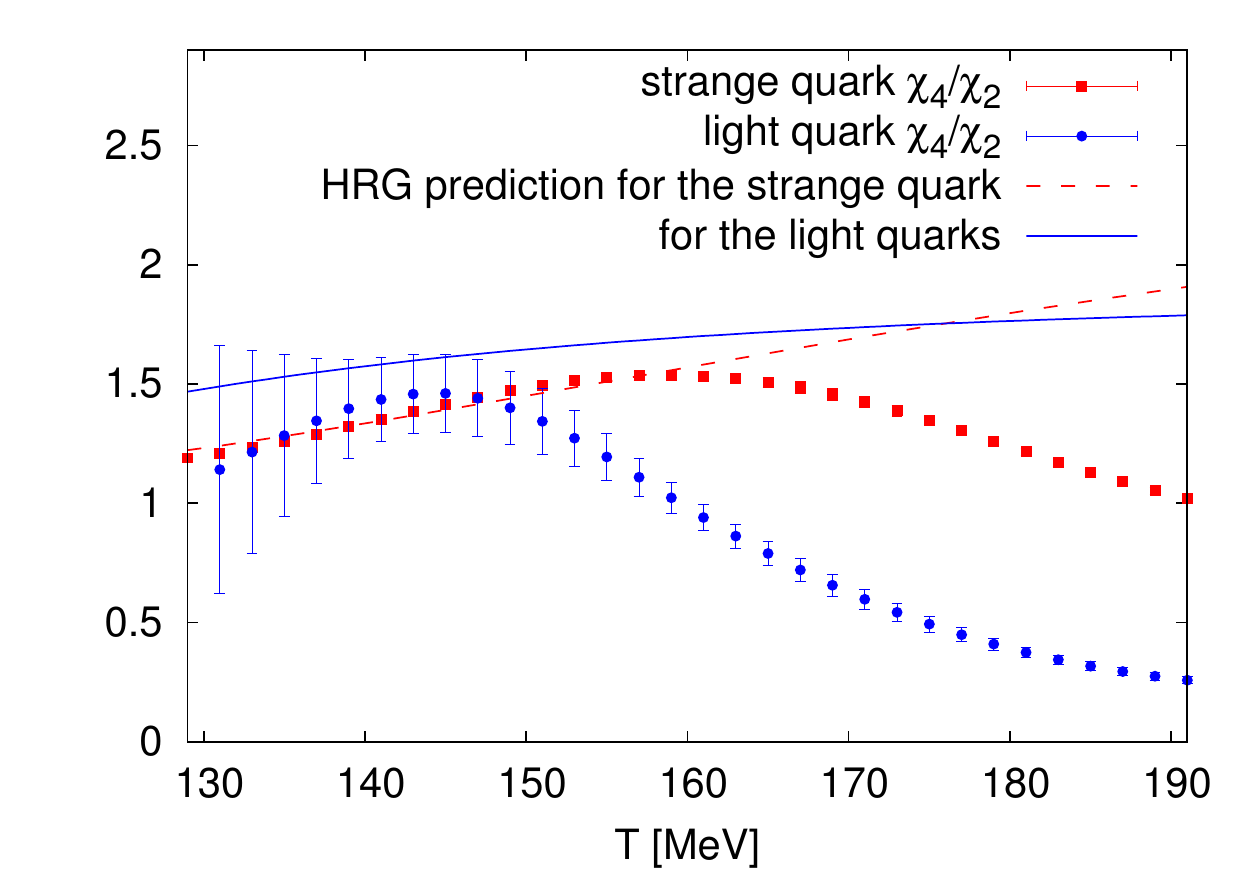}
}
\caption{\label{c24} The $T$-dependence of the $\chi_4/\chi_2$ ratio for light and strange quarks in the continuum limit. The lattice data are compared to HRG calculations.
}
\end{figure}

For the sake of defining observables which can, in principle, be measured in
experiments, we focus on more basic susceptibility combinations.
The most attractive quantity  to the purpose is $\chi_4^f/\chi_2^f$, since the
ratio does not depend on the volume.  Similar ratios have been proposed to
determine the chemical freeze-out temperature independent of any statistical
model assumptions \cite{Karsch:2012wm,Bazavov:2012vg,Borsanyi:2013hza}. Its
non-monotic behavior as a function of the temperature has also been suggested
as an indicator for the deconfinement transition \cite{Ejiri:2005wq}.
Fig.~\ref{c24} shows the $T$-dependence of $\chi_4/\chi_2$ for light and
strange quarks. For the light quark susceptibilities (thus for observables
related to net up+down quark numbers) the pion contribution, which is
notoriously difficult to calculate on the lattice, is absent by
definition.  The figure shows two characteristic features: a) each lattice
calculation exhibits a kink (or peak) at a particular temperature and b)
this kink coincides with the temperature at which the lattice curve starts to
deviate from the HRG predictions.  Interestingly enough, the separation between
the kinks of the two flavors corresponds to the previously mentioned
$\approx$15~MeV.  In a scenario, in which the highest temperature where the HRG
and lattice QCD agree is indicating a ``deconfinement" or ``liberation"
temperature for a particular flavor Fig.~\ref{c24} further supports the flavor
separation of the characteristic temperatures.

The presented lattice results show that these quantities, if measured with high
accuracy, are good thermometers. The challenge is to unambiguously demonstrate
such a flavor hierarchy experimentally. The effect could manifest itself to
first order in the multiplicity distribution of identified particles. In case
strange hadrons form at a higher temperature than their light quark
counterparts, their abundance will be enhanced relative to a common low
temperature freeze-out scenario. Therefore, the comparison of measured yields
to a statistical hadronization model enables one to determine chemical
freeze-out temperatures (T$_{ch}$) in a flavor-separated way.  First results
from ALICE indicate that the T$_{ch}$ of strange hadrons is about 16 MeV higher
than that of light hadrons (164 vs 148 MeV)
\cite{Preghenella:2011np,Abelev:2012wca}. 
As in the case of the lattice parameters, this sensitivity to the freeze-out temperature, extracted from a statistical hadronization fit, is most pronounced for the multi-strange baryons.
These temperature fits are
model-dependent, though, and a direct comparison to the temperatures extracted
from quark susceptibilities in lattice QCD likely requires corrections. For
example, it was suggested that final state interactions between hadrons might
modify the baryon yields \cite{Steinheimer:2012rd,Becattini:2012xb}.

A more precise verification, less prone to alternate explanations, can be
obtained by using a higher order moment analysis of particle identified yields,
since those moments can be directly related to the higher order
susceptibilities on the lattice \cite{Karsch:2012wm}. In particular, the
product of kurtosis and the square of the variance of the net multiplicity
distributions corresponds to the susceptibility ratio shown in Fig. \ref{c24}.
Even for this analysis, though, the caveat is the exact relation between the
quark-based observable and the hadron-based measurement.  Specifically, one
needs to determine how many hadron species have to be measured in order to
fully capture the transition behavior of the respective quark flavor.
Preliminary studies show that taking into account only the dominant mesonic
states is not sufficient. Baryonic states require significant acceptance and
efficiency corrections, though, even under the assumption of statistical
behavior of the higher moments of reconstructed particles in the detector
acceptance \cite{Bzdak:2012ab}.  Furthermore, corrections for hadronic
resonance decays have to be considered. Still, based on first correction
studies and initial results from the RHIC experiments for a subset of
identified particle species \cite{McDonald:2012ts}, we expect that the flavor
separation, suggested by several characteristic temperatures obtained on the
lattice, can be experimentally verified or falsified for the freeze-out
temperatures, too. Once the experimental results are approaching a certain
value with small errors, one can extend the studies shown in Fig. \ref{c24} in
this particular temperature region with additional higher precision lattice
data.


In conclusion we have presented, for the first time, high precision continuum
extrapolated lattice calculations of flavor-specific higher order
susceptibility combinations at zero baryo-chemical potential and high
temperatures. We have shown that flavor-dependent patterns emerge in the
crossover region of the QCD transition.
The $T$-dependence of the
examined observables hints at a flavor separation,
especially for quantities which are most sensitive to multi-strange states.
This flavor separation is an
obvious consequence of the mass difference between the light and
strange quarks. These subtle differences of about 15~MeV in the
characteristic temperatures are only visible thanks to the large accuracy of the
latest lattice calculations. We have proposed an experimental program that
might allow an observation of a possible similar separation in the
freeze-out temperatures. One can look for these effects at LHC, and
potentially RHIC energies if the net multiplicity distributions of identified
particles in the relevant quark sectors can be measured and efficiency
corrected to the same high accuracy as the lattice QCD calculations.

\textrm{Acknowledgments:}
This project was funded by the DFG grant SFB/TR55. The work of R. Bellwied is
supported through DOE grant DE-FG02-07ER41521. The work of C. Ratti is
supported by funds provided by the Italian Ministry of Education, Universities
and Research under the Firb Research Grant RBFR0814TT. S. D. Katz is funded by
the ERC grant ((FP7/2007-2013)/ERC No 208740) and the Lend\"ulet program of the
Hungarian Academy of Sciences ((LP2012-44/2012).  The simulations were
performed on the QPACE machine, the GPU cluster at the Wuppertal University and
on JUQUEEN (the Blue Gene/Q system of the Forschungszentrum Juelich).

\end{document}